\documentclass[3p,times]{elsarticle}
\usepackage{ecrc}
\author{Yetkin Yilmaz on behalf of the CMS Collaboration}

\volume{00}
\firstpage{1}
\journalname{Nuclear Physics A}
\runauth{}
\jid{nupha}

\usepackage{amssymb,epsfig,amsmath}

\usepackage[figuresright]{rotating}

\usepackage{setspace}

\def\be{\begin{equation}}
\def\ee{\end{equation}}
\def\bea{\begin{eqnarray}}
\def\eea{\end{eqnarray}}

\def\npart{$N_\text{part}$}
\def\pt{$p_\text{T}$}

\def\aptr{\ensuremath{\left<p_\text{T,2}/p_\text{T,1}\right>}}

\def\dphi     {\ensuremath{\Delta\phi_{1,2}}}

\def\npart    {\ensuremath{N_\text{part}}}

\def\PbPb  {\mbox{PbPb}}

\def\dphi {\ensuremath{\Delta\phi_{1,2}}}
\def\GeVc {GeV/$c$}

\begin{document}
\pagenumbering{gobble}

\begin{figure}[h!]
\begin{center}
\includegraphics[width=1.\textwidth]{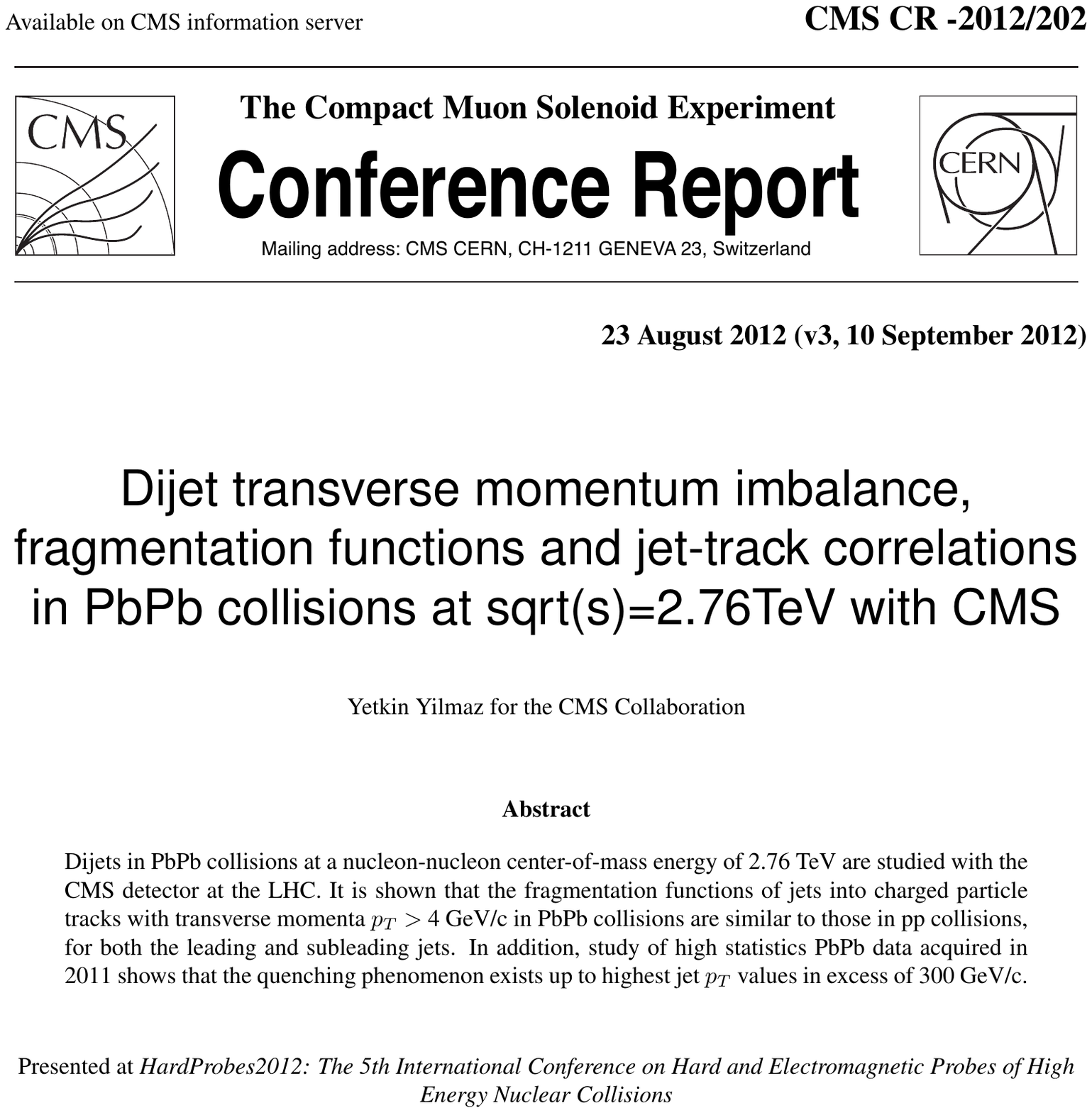}
\end{center}
\end{figure}

\clearpage

\setcounter{page}{1}
\pagenumbering{arabic}

\begin{frontmatter}



\title{Dijet transverse momentum imbalance, fragmentation functions and jet-track correlations in PbPb collisions at $\sqrt{s_{NN}}=2.76$ TeV with CMS}


\address{Massachusetts Institute of Technology, Cambridge, MA 02139, USA}

\begin{abstract}
Dijets in PbPb collisions at a nucleon-nucleon center-of-mass energy of 2.76 TeV are studied with the CMS detector at the LHC. It is shown that the fragmentation functions of jets into charged particle tracks with transverse momenta \pt\ $>$ 4 GeV/c in PbPb collisions are similar to those in pp collisions, for both the leading and subleading jets. In addition, study of high statistics PbPb data acquired in 2011 shows that the quenching phenomenon exists up to highest jet \pt\ values in excess of 300 GeV/c.
\end{abstract}


\end{frontmatter}


\begin{figure}[h!]
\begin{center}
\includegraphics[width=1.\textwidth]{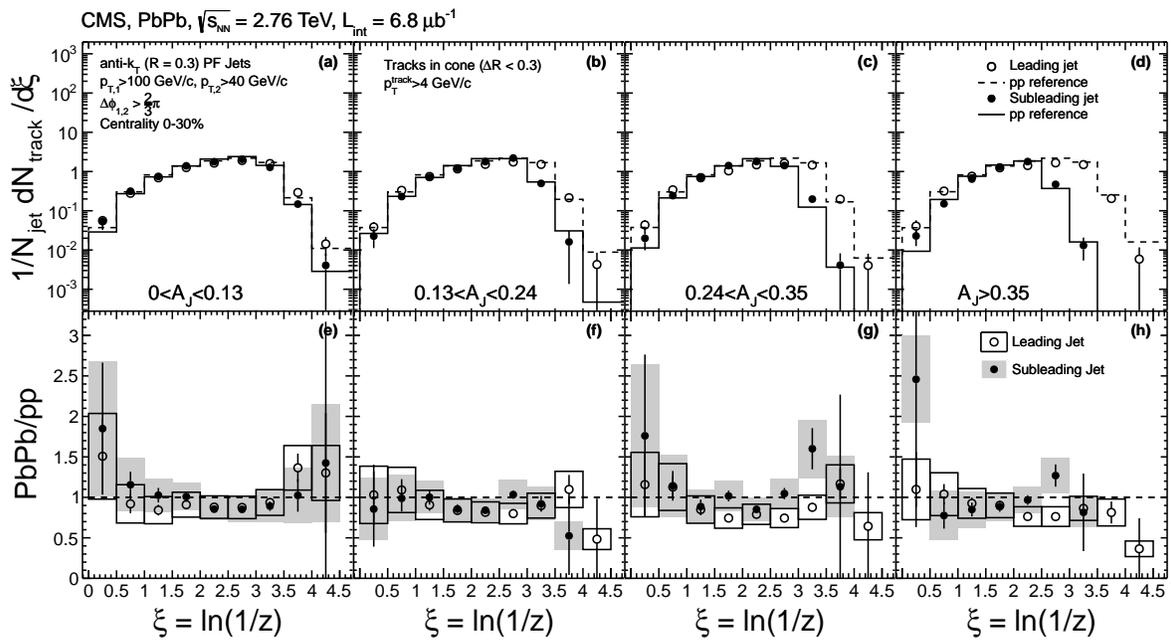}
\end{center}
\caption{The top row shows the fragmentation functions in PbPb data and pp-based reference for various dijet asymmetry selections. The bottom row shows the ratio of each fragmentation function to its pp-based reference. The error bars represent the statistical uncertainty and the boxes represent the systematic uncertainty.}
\label{fig:FFAJ}
\end{figure}

The energy loss phenomenon in the medium that is formed in highly energetic PbPb collisions, which was previously observed in hadron measurements at RHIC, is explored in more detail with fully reconstructed jet measurements that are performed in LHC experiments~\cite{Aad:2010bu,Chatrchyan:2011sx}. This paper discusses the recent studies from CMS~\cite{CMS} on jet fragmentation and the momentum dependence of the quenching.

\section{Fragmentation functions}

Earlier results from CMS~\cite{Chatrchyan:2011sx}, with the dataset of the 2010 LHC run with PbPb ions, have revealed various aspects of the energy loss mechanism. Although an angular de-correlation between the jets is not observed, the average transverse momentum imbalance is observed to be increasing with collision centrality, which is attributed to energy loss in medium. 
The properties of jets are further investigated qualitatively by the study of fragmentation functions~\cite{FragmentationPAS} , where tracks with \pt $>$ 4 \GeVc\ within a cone of $\Delta R<0.3$ are correlated to the jet, where $\Delta R = \sqrt{(\Delta \eta)^{2} + (\Delta \phi)^{2}}$ between the track and the jet. The momenta of each track is projected onto the jet axis in the reference frame where the two jets have opposite pseudorapidity. The fragmentation functions are plotted as a function of $\xi = \ln (1/z)$ where $ z = p_{\parallel}^{\mathrm{track}}/p^{\mathrm{jet}}$, $p_{\parallel}^{\mathrm{track}}$ is the momentum component of the track along the jet axis, and $p_{\mathrm{jet}}$ is the magnitude of the jet momentum.

The distribution of $\xi$ is shown in Fig.~\ref{fig:FFAJ} in bins of dijet asymmetry, $A_{J} = (p_{\mathrm{T},1}-p_{\mathrm{T},2})/(p_{\mathrm{T},1}+p_{\mathrm{T},2})$.
In any given $A_{J}$ selection, fragmentation of jets display the same pattern in PbPb collisions and in pp collisions.

\clearpage

\section{Dijet imbalance as a function of leading jet \pt}

With the availability of a large dataset from the 2011 LHC run with PbPb ions, the dijet imbalance is investigated through a more differential approach~\cite{Chatrchyan:2012nia}.

Dijets with \dphi $>$ $2\pi/3$ are selected.
The contamination from fake jets, due to background fluctuations, is subtracted as estimated from dijet events with \dphi $<$ $\pi/3$.

\begin{figure}[h!]
\begin{center}
\includegraphics[width=.8\textwidth]{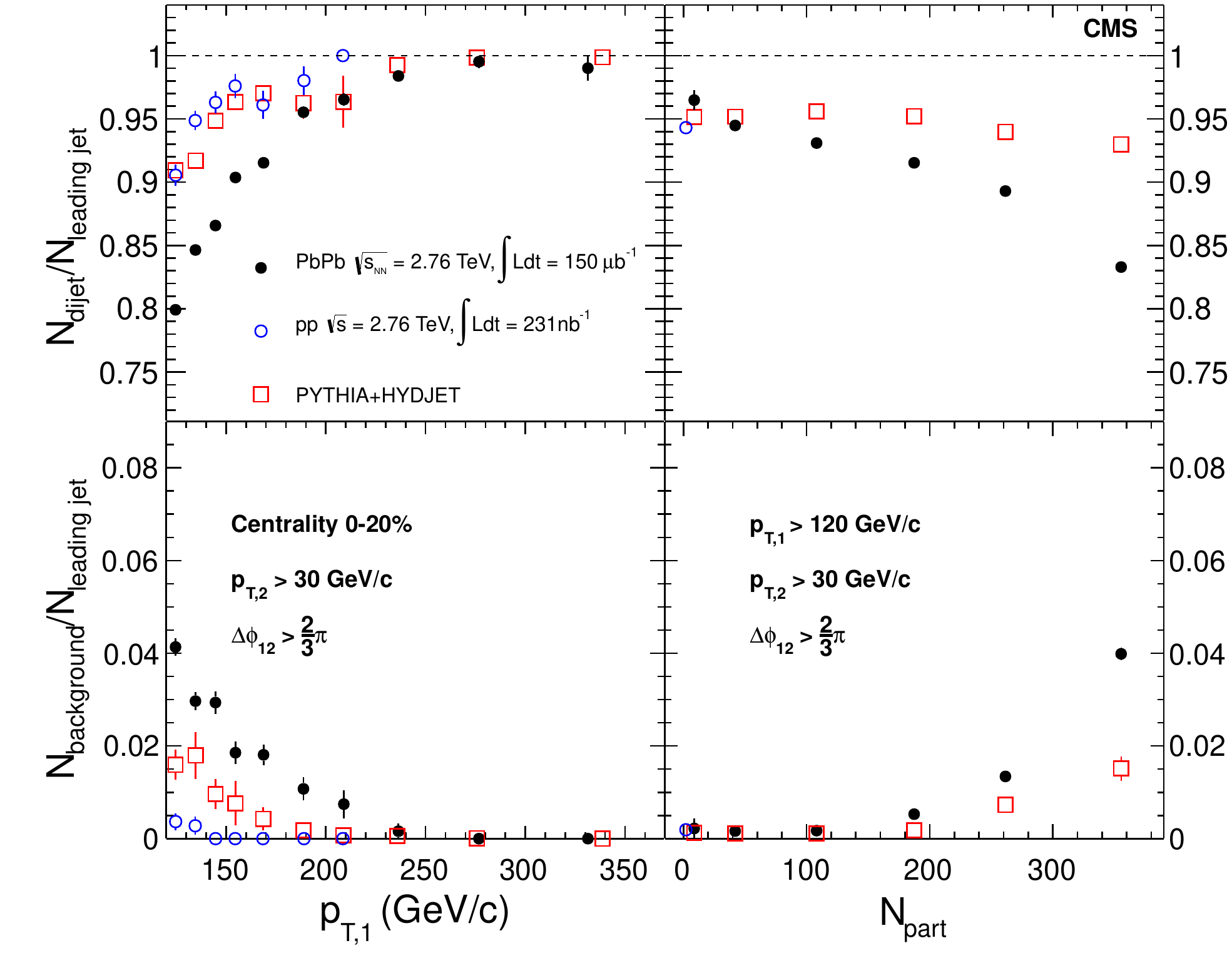}
\label{SubRate}
\caption{
Fraction of events with a genuine subleading jet with $\dphi > 2\pi/3$, as a function of leading jet $p_{\text{T},1}$ (left) 
and \npart\ (right). The background due to underlying event
fluctuations is estimated from $\dphi < \pi/3$ events and subtracted from the number of dijets. 
The fraction of the estimated background is shown in the bottom panels.
The error bars represent the statistical uncertainties.}
\end{center}
\end{figure}

It is observed that when the leading jet of the event has a \pt\ higher than 180 GeV/c, it is more than 95\% of the time accompanied by a recoiled partner in the opposite direction in azimuth. The fraction of such correlated events after background subtraction and the fraction of the estimated background are shown in Fig.~\ref{SubRate} as a function of leading jet \pt\ and event centrality.

\begin{figure}[h!]
\begin{center}
\includegraphics[width=1.0\textwidth]{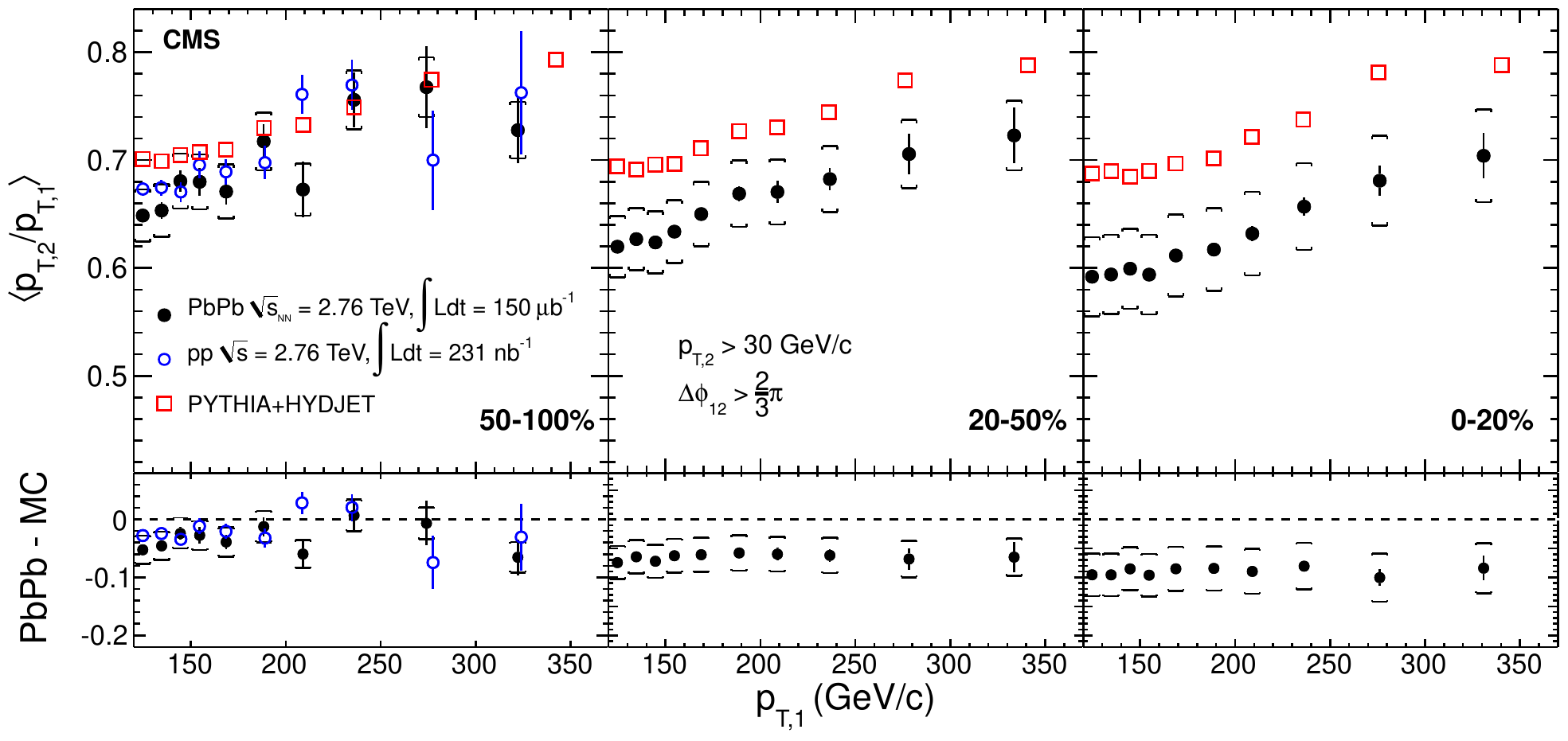}
\caption{
Average dijet momentum ratio $p_{\text{T},2} / p_{\text{T},1}$ as a function of 
leading jet \pt\ for three bins of collision centrality, from peripheral to central collisions,
corresponding to selections of 50--100\%,  30--50\% and 0--20\%  of the total inelastic cross section.
Results for \PbPb\ data are shown as points with vertical bars and brackets indicating 
the statistical and systematic uncertainties, respectively.  Results for {\sc{pythia+hydjet}} are shown as squares. In the 50--100\% centrality bin,
results are also compared with pp data, which is shown as the open circles.
The difference between the \PbPb\ measurement and the {\sc{pythia+hydjet}} expectations is shown in the bottom panels. }
\label{deltaPt}
\end{center}
\end{figure}

In Fig.~\ref{deltaPt}, the average ratio of subleading jet \pt\ to the leading jet \pt, \aptr, is shown as a function of leading jet \pt\ in different bins of centrality. In the central events, a significant shift of the \aptr\ with respect to the MC and pp results is observed. This shift, while changing monotonically with centrality, does not show a significant dependence on the leading jet \pt. Since both data and MC include an intrinsic imbalance from hard gluon radiation and detector resolution, the implications on the absolute amount of energy loss should be extracted via realistic models of quenching which take into account these effects.

\end{document}